\documentclass[10pt,letterpaper]{article}
\pdfoutput=1
\usepackage[a4paper,
            bindingoffset=0.2in,
            left=1in,
            right=1in,
            top=1in,
            bottom=1in,
            footskip=.25in]{geometry}
\usepackage{tabularx,booktabs}
\usepackage{url}
\usepackage{fullpage}
\usepackage{graphicx}
\usepackage{lipsum}
\newcommand{\junk}[1]{}

\usepackage{algorithm}
\usepackage{algorithmic}
\usepackage{amsmath}
\usepackage{amsfonts}

\usepackage{subfigure}
\usepackage{float}

\usepackage{multirow}
\usepackage{setspace}
\usepackage{mathrsfs}
\usepackage{authblk}
\usepackage{times}
\usepackage{subfiles}

\begin{document}
\thispagestyle{empty}

\title{Contrastive Cycle Adversarial Autoencoders for Single-cell Multi-omics Alignment and Integration}

\author[1,4]{Xuesong Wang}
\author[1]{Zhihang Hu}
\author[2,3]{Tingyang Yu}
\author[1]{Ruijie Wang}
\author[1]{Yumeng Wei}
\author[5]{Juan Shu}
\author[6]{Jianzhu Ma}
\author[1,4]{Yu Li \thanks{Corresponding Author. Email: liyu@cse.cuhk.edu.hk}}
\affil[1]{\small Department of Computer Science and Engineering, CUHK, Hong Kong SAR, China}
\affil[2]{\small Department of Mathematics, CUHK, Hong Kong SAR, China}
\affil[3]{\small Department of Information Engineering, CUHK, Hong Kong SAR, China}
\affil[4]{\small The CUHK Shenzhen Research Institute, Hi-Tech Park, Nanshan, Shenzhen, 518057, China}
\affil[5]{\small Purdue University, West Lafayette, IN 47907, United States}
\affil[6]{\small Institute for Artificial Intelligence, Peking University, Beijing, 100871, China}

\date{}

\maketitle
\begin{abstract}
Muilti-modality data are ubiquitous in biology, especially that we have entered the multi-omics era, when we can measure the same biological object (cell) from different aspects (omics) to provide a more comprehensive insight into the cellular system. 
When dealing with such multi-omics data, the first step is to determine the correspondence among different modalities. In other words, we should match data from different spaces corresponding to the same object. This problem is particularly challenging in the single-cell multi-omics scenario because such data are very sparse with extremely high dimensions. Secondly, matched single-cell multi-omics data are rare and hard to collect. Furthermore, due to the limitations of the experimental environment, the data are usually highly noisy. To promote the single-cell multi-omics research, we overcome the above challenges, proposing a novel framework to align and integrate single-cell RNA-seq data and single-cell ATAC-seq data. Our approach can efficiently map the above data with high sparsity and noise from different spaces to a low-dimensional manifold in a unified space, making the downstream alignment and integration straightforward. Compared with the other state-of-the-art methods, our method performs better in both simulated and real single-cell data. The proposed method is helpful for the single-cell multi-omics research. The improvement for integration on the simulated data is significant.\\
\textbf{Keywords}: Single-cell multi-omics, Multi-omics integration, Cycle autoencoders, Contrastive learning, scRNA-seq and scATAC-seq
\end{abstract}

\clearpage

\section{Introduction}
\begin{figure}[!t]
    \centering  
    \subfigure[Problem setting]{
    
    \includegraphics[width=0.275\textwidth]{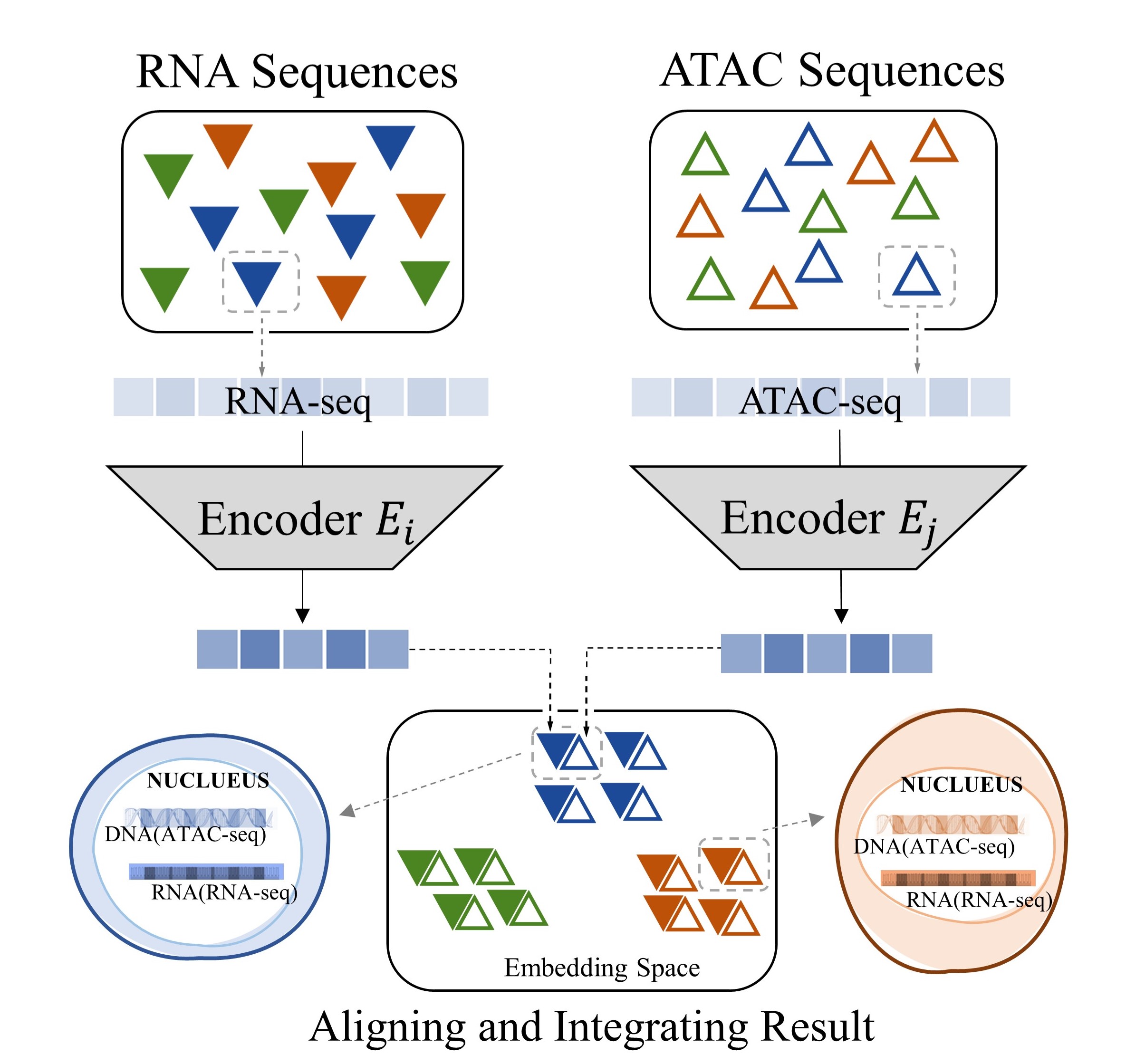}}
    \subfigure[Framework]{
    \includegraphics[width=0.7\textwidth]{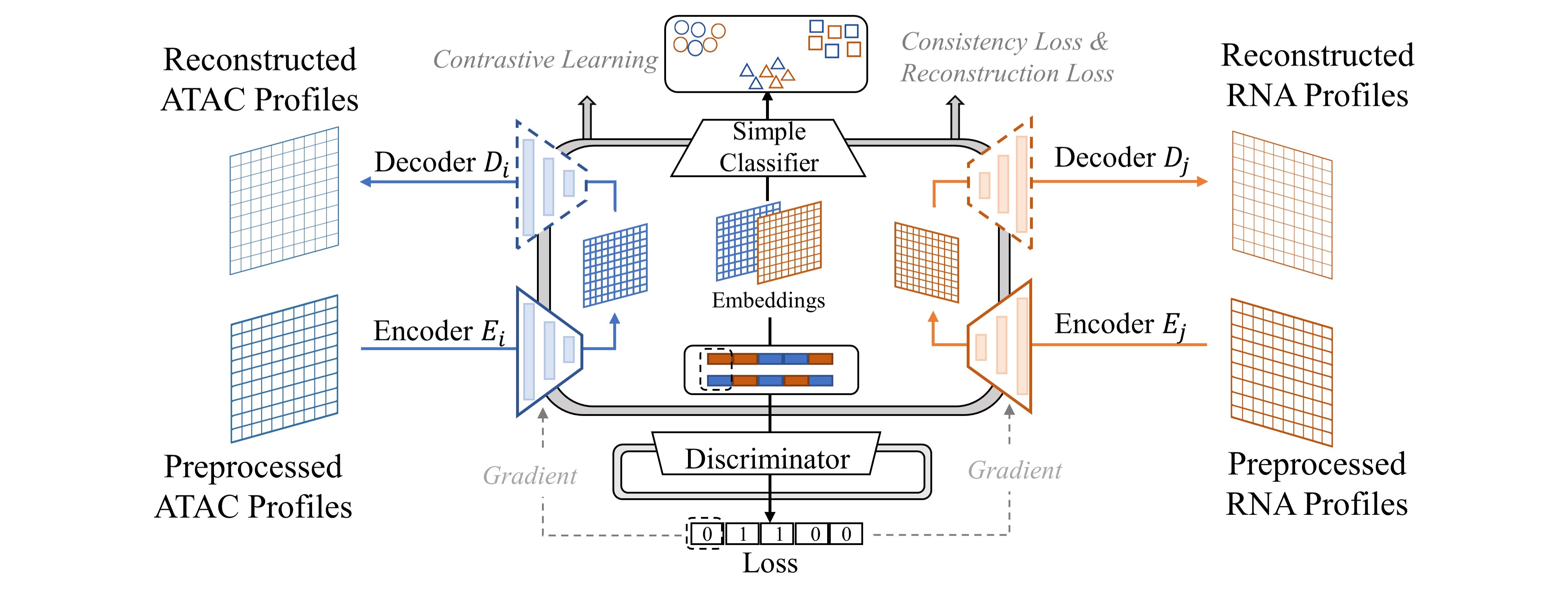}}
    
    \caption{\textbf{(a)} scRNA-seq and scATAT-seq are two sequencing technologies, which measure different aspects of the same cell. The two experiments are usually done at different time points. We aim at identifying the correspondence between the two kinds of data from the same set of cells. \textbf{(b)} The Con-AAE framework uses two autoencoders to map the two kinds of sequence data into two low dimensional manifolds, forcing the two spaces as unified as possible with the adversarial loss and latent cycle-consistency loss. For the alignment task, we train the models without pairwise information but consider the data noise explicitly by utilizing self-supervised contrastive learning. For the integration task, we train the framework with labeled data. }
    \label{fig:Problem&framework}
\end{figure}

  Multi-modality data, which describe an object from different perspectives, help us understand the world more comprehensively \cite{baltruvsaitis2018multimodal}. Such data are typical and popular in biology, which are referred as multi-omics data. That is, we can measure a cell from different aspects (omics) to provide a more comprehensive insight into the cellular system. To achieve that, we should obtain multi-omics data related to the same cell, which is not a trivial task.
Although multi-omics profiling approaches for the same set of single cells have become available \cite{gala2021consistent}, such as single-cell RNA sequencing (scRNA-seq) and single-cell Assay for Transposase Accessible Chromatin sequencing (scATAC-seq), which describe the same cell from different perspectives, the experiments are usually done at different time points for a set of cells. Consequently, we have the multi-omics data for a group of cells, but the correspondence between different modalities for a single cell is missing (Figure \ref{fig:Problem&framework}a).  More specifically, we want to obtain the high-throughput paired multi-omics data for every single cell, which is referred as \textit{alignment}. On the other hand, even for data within the same modality, the data distribution can be inconsistent because of the subtle differences in measurement processes \cite{cao2018joint}, such as measurement time or equipment used, which is referred as \textit{batch effects}. They should also be considered when we are studying the correspondence among different modalities. Considering such batch effects, we need to integrate different multi-omics data from the same batch, which is called \textit{integration}. The above tasks of alignment and integration are very useful and interesting, but difficult, considering the distribution shifting within and across different modalities and the sparsity and high dimension of the single-cell data \cite{klein2015droplet,macosko2015highly,buenrostro2015single}.

Some computational methods have been proposed to deal with this important but difficult problem, aligning and integrating data from different omics. 
However, the most popular works in the single-cell field focus on integrating datasets of the same modality \cite{butler2018integrating,trong2020semisupervised,stuart2019comprehensive}, which are not suitable for different modalities.
People usually integrate and align multi-omics data in the learned low-dimensional embedding space using dimension reduction techniques, such as Principal Component Analysis (PCA) \cite{bersanelli2016methods,argelaguet2018multi} and nonlinear successors of the classic Canonical Correlation Analysis (CCA) \cite{stanley2020harmonic}. The typical examples are Seurat \cite{stuart2019comprehensive} and Deep Classic Canonical Correlation (DCCA) \cite{andrew2013deep}. Seurat relies on the linear mapping of PCA and aligns the manifold vector based on linear methods Mutual Nearest Neighbors (MNNs) and CCA, which weaken its ability to handle nonlinear geometrical transformations across cellular modalities \cite{cao2020manifold}. DCCA can be effective for nonlinear transformation benefiting from deep learning, but according to the results of our experiments, it is not robust enough when the signal-to-noise ratio (SNR) is low. We also tried Maximum Mean Discrepancy (MMD) \cite{borgwardt2006integrating} replacing CCA in the embedding space, but the performance is also not good enough. 
Several methods requiring no correspondence information were derived under various advanced machine learning techniques, such as Pamona \cite{cao2020manifold}, MATCHER \cite{welch2017matcher},  MMD-MA \cite{singh2020unsupervised}, UnionCom \cite{cao2020unsupervised}, SCOT \cite{demetci2020gromov}. Although these methods are unsupervised and achieve integrative performance with encouraging results \cite{demetci2020gromov}, there are still other additional conditions required. For examples, MMD-MA and UnionCom, Pamona need the user specify several hyperparameters, while MATCHER and SCOT require the assumption that all datasets share the same underlying structure across cellular modalities \cite{cao2020manifold}. Selecting hyperparameter values can be difficult and may need prior information and such assumption can be ineffective confronting dataset-specific cell types/structures across the single-cell datasets \cite{cao2020manifold}.
Deep learning methods are promising to provide alignment and transfer learning between datasets \cite{li2019deep,li2020modern}.
Deep generative models, such as cycleGANs \cite{zhu2017unpaired}, MAGAN \cite{wang2017magan}, RadialGAN \cite{yoon2018radialgan} and starGAN \cite{choi2018stargan}, are used to learn a nonlinear mapping from one domain to another in unsupervised way. 
But the above transitions are almost within the same modality and can be disturbed by noise or sparsity in the data \cite{stanley2020harmonic}. The scenario of multi-omics translation and alignment is much more complicated. 
Some other works indeed propose methods to align multi-omics data with multiple autoencoders  \cite{dai2021multi,zhang2019integrated,ma2019integrate}. However, such methods also can be seriously affected by noise or sparsity, which is a fundamental characteristic of single-cell data.

In general, there are four major challenges in multi-omics alignment. Firstly, although we have a large amount of unaligned multi-omics data, the aligned data is very scarce. Secondly, the single-cell omics data is of very high dimension and extremely sparse. For example, the unprocessed scRNA-seq data contain around 25,000 genes (features) while, in the data matrix, 40\% to 80\% of the values are 0. Furthermore, as we have discussed, the data are highly noisy \cite{klein2015droplet,macosko2015highly,buenrostro2015single}. Finally, although both scRNA-seq data and scATAC-seq data describe the cell status, they contain very different information, and the mapping between the two kinds of data is highly complicated.
To promote the single-cell multi-omics data analysis, we propose a framework based on Contrastive cycle adversarial Autoencoders (Con-AAE), which can resolve the above challenges precisely (Figure \ref{fig:Problem&framework}b).
Con-AAE uses two autoencoders to map the two modal data into two low-dimensional manifolds, forcing the two spaces as unified as possible. We use two novel loss terms to achieve that. The first term is called adversarial loss. That is, we combine GAN with autoencoders, forcing the two autoencoders to produce unified embedding to deceive the discriminator, which is designed to distinguish whether two embedding factors are from the same modality or not. However, only using the adversarial loss may lead to model collapse. To avoid the problem, we further propose a novel latent cycle-consistency loss. In a nutshell, the embedding produced by the scRNA-seq encoder will go through the scATAC-seq decoder and encoder to produce another cycled embedding, for which we can check the consistency between the original embedding the cycled embedding. In addition to the above two loss terms, for the alignment task, we train the models without pairwise information but consider the data noise explicitly by taking advantage of self-supervised contrastive learning. For the integration task, we train the framework with labeled data. 
We extensively perform experiments on two real-world datasets and a group of simulated datasets. The comprehensive experiments on both the simulated datasets and real-world datasets show that our method has better performance and is more robust than the other state-of-art methods.

\section{Methods}

In this section, We give our framework in details below with Fig. \ref{fig:Problem&framework}b which illustrates the whole pipeline. To start with, we formalize the alignment problem as,

\begin{eqnarray}
\forall (r,a) \in \mathscr{T}, f(a) = h(r).
\end{eqnarray}

We denote (r,a) as a pair of RNA-seq and scATAC-seq taken from the same cell. We would like to find two mappings $f$ and $h$ such that for any aligned \{r,a\} pairs in $\mathscr{T}$, $f$ would maps the scRNA-seq profile and scATAC-seq profile to a shared embedding space. Due to the limitations of available real-world aligned data, we are actually working on an unsupervised problem and the results are evaluated on a few available aligned pairs.

The integration problem could be justified as a classification problem, and the objective results to finding the corresponding batch of each scRNA-seq or scATAC-seq profile. The groundtruth labels are available, and, therefore, it can be trained in a supervised way. For $\forall x \in$\{scRNA-seq\}$\cup$\{scATAC-seq\}, we want to train a classifier $g$ such that

\begin{equation}
g(x) = label(x).
\end{equation}

Our main model is built upon the framework of Adversarial 
Auto-Encoders \cite{makhzani2015adversarial} specialized for modality task, integrating our novel embedding consistency module and contrastive training process \cite{chen2020simple}. The intuition behind is that multi-omics from a single-cell data should obtain commonality, and their mappings could live in a unified low-dimensional manifold, which therefore makes alignment and integration task more accurate.

\subsection{Adversarial Auto-Encoders}

The usage of adversarial auto-encoders aims to map different omics into a unified latent manifold while able to reconstruct these different aspects. Therefore, as shown in Figure \ref{fig:Problem&framework}b, we are using a coupled set of encoders $\{E_{i},E_{j}\} $\cite{hira2021integrated} to map \{scATACs-eq,scRNA-seq\} into manifolds $\{Z_{i},Z_{j}\}$, and decoders $\{D_{i},D_{j}\}$ could decode the embedded manifolds back to the original distribution. The reconstruction loss is defined as follows,

\begin{eqnarray}
    L_{recon}&=&\mathbb{E}_{x \sim p_{rna}}d(x,D_{j}(E_{j}(x)))  \nonumber  \\
     &+ &\mathbb{E}_{x \sim p_{atac}}d(x,D_{i}(E_{i}(x))),
\end{eqnarray}

whereas $d$ stands for indicated distance in the embedding space. Discriminator $\mathscr{D}$ tries to align these embedded manifolds and works in the sense that input $x\in Z_{j}$, $\mathscr{D}(x) = 1$ or $x\in Z_{i}$, $\mathscr{D}(x) = 0$. 

\begin{eqnarray}
    L_{adv}&=&\mathbb{E}_{x \sim p_{rna}}[\log\mathscr{D}(E_{j}(x))]  \nonumber  \\
     &+ &\mathbb{E}_{x \sim p_{atac}}[\log(1-\mathscr{D}(E_{i}(x)))].
\end{eqnarray}

The above losses $L_{recon}$ and $L_{adv}$ are trained together with the same weights. 

\subsection{Latent Cycle-Consistency Loss}
The backbone framework enforces the embedding manifolds to gradually align. 
However, a critical problem underlying is that since RNA and ATAC data are sparse in a high dimensional domain, the training procedure above only aligns and trains on those regions where the data exist.

For instance, if a region $A$ in the embedding space around $E_{j}(x'),x'\in$\{scRNA-seq\} does not involve any existing $E_{i}(x),x\in$\{scATAC-seq\}. Then, neither the decoder $D_{i}$ nor the encoder $E_{i}$ is trained on $A$, thus they would not compute in a ``reverse'' mapping way, and the result of $E_{i}D_{i}E_{j}(x)$ would be unreasonable or may not lie on the aligned manifold. 
This critical problem causes the difficulty of inferring from scRNA-seq profile to scATAC-seq profile directly.

\begin{figure}
    \centering  
    \subfigure[Illustration of latent cycle-consistency loss.]{
    
    \includegraphics[width=0.48\textwidth]{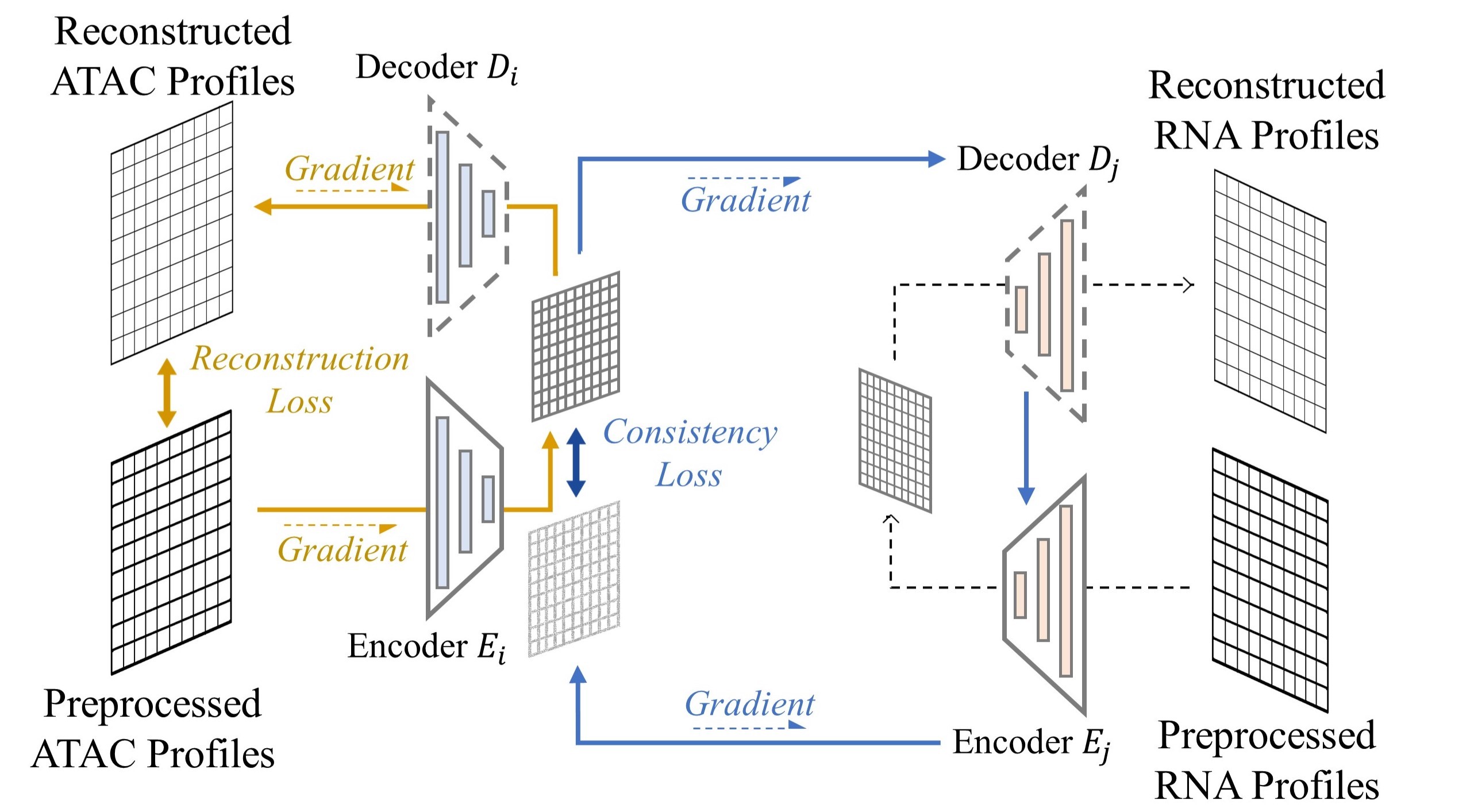}}
    \subfigure[Illustration of latent contrastive loss.]{
    \includegraphics[width=0.48\textwidth]{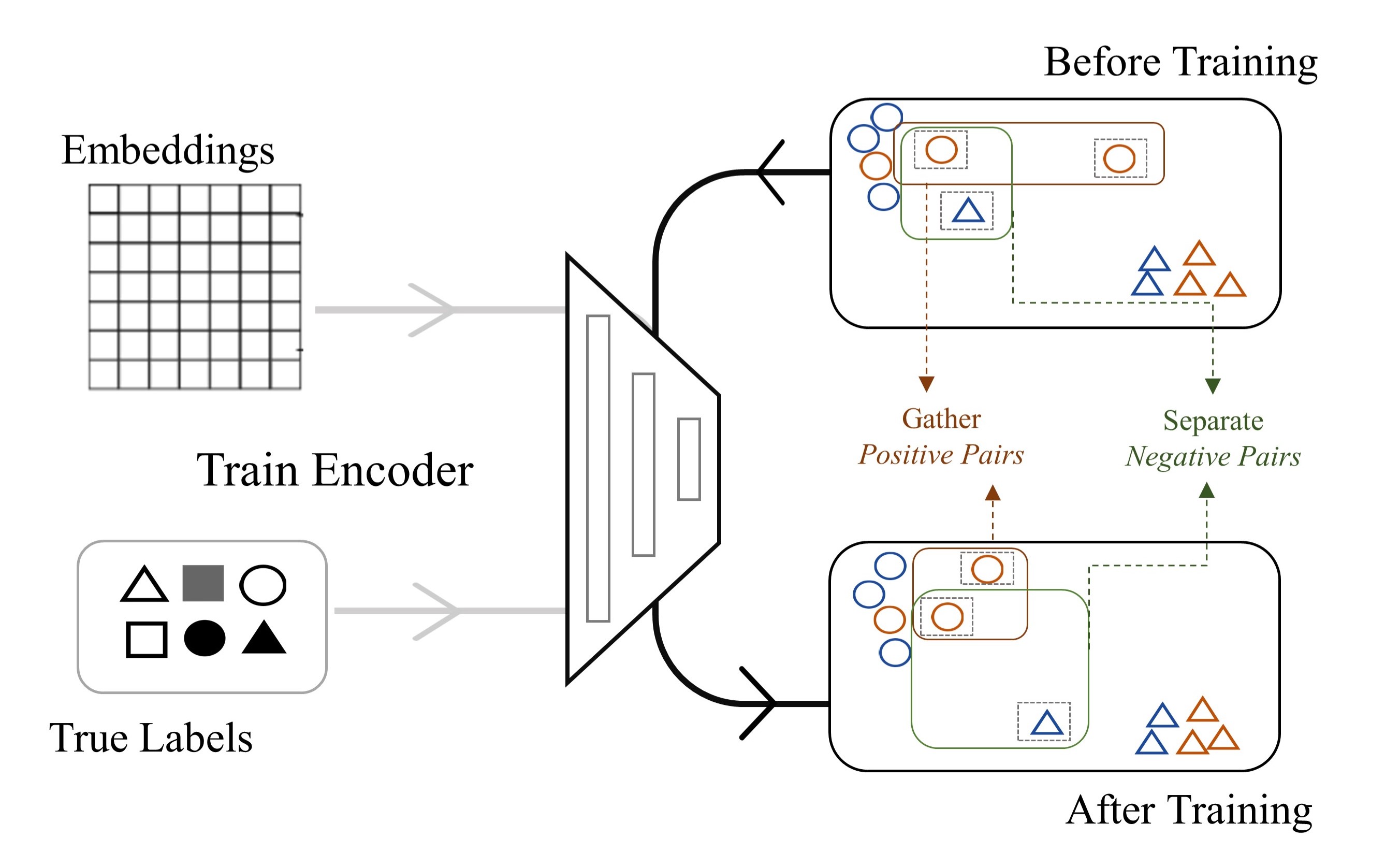}}
    
    \caption{\textbf{(a)} The embedding produced by the first encoder will go through the second decoder and encoder to produce another cycled embedding, for which we can check the consistency between the original embedding the cycled embedding. \textbf{(b)} The contrastive loss minimizes the distance between positive pairs and maximizes the distance between negative pairs. This loss makes our method more robust to noise.}
    \label{fig:cycle&contrastive}
\end{figure}

Therefore, we introduce a latent consistency loss shown in Figure \ref{fig:cycle&contrastive}a \cite{zhu2017unpaired}\cite{DBLP:conf/ijcai/HuW19} to resolve this problem,

\begin{eqnarray}
    L_{cyc}&=&\mathbb{E}_{x \sim p_{rna}}[d(E_{j}(x),E_{i}D_{i}E_{j}(x))]  \nonumber  \\
     &+ &\mathbb{E}_{x \sim p_{atac}}[d(E_{i}(x),E_{j}D_{j}E_{i}(x))].
\end{eqnarray}

$L_{cyc}$ aims to train the set of encoder-decoder on the domain of where different omics data may not exists, which enforces the smoothness and consistency on those regions. In this way, we could compare the embedding of $E_{j}(x),x\in$\{scATAC-seq\} directly with the existing scRNA-seq embedding around it.

\subsection{Supervised Contrastive Loss}

The above framework works in an unsupervised manner such that the embedded latent manifolds of multi-omics align properly. A simple classifier trained on our latent space could achieve rather good accuracy on the integration task.

We could further improve our work on both tasks taking advantage of the ground-truth cell type labels. The cell type label could refer to biological cell types, or the label of data batches collected from different time or platform. Following the idea of contrastive learning \cite{schroff2015facenet,han2021self}, we employ a contrastive loss in embedding space to the training procedure. It enforces smaller In-Batch distance and larger Between-Batch distance. In-Batch refers to different modalities data collected from the same batch, vice versa. We equally treat both modalities in contrastive training, which benefits the alignment task in the sense that multi-omics of the same single-cell data should obviously belong to the same batch. We show that lowering the In-Batch distance indeed improves the alignment accuracy in the below ablation studies. On the other hand, contrastive training benefits integration task by enabling the decision boundary smoother and more robust.

In practice, we first encode data from two modalities to the embedding space. Define the embedding by $z\in\mathbb{Z}$. Given $z^{a}$ as anchor vector in latent space, we select an $z^{p}$ such that $argmax_{z^{p}}\{ d(z^a,z^p)\},label\{z^a\} = label\{z^p\}$, which is named hard positive. The intuition of hard positive is to find a vector furthest from the anchor 
within same cluster. Similarly, we have $z^n$ as hard negative such that $argmin_{z^{n}} \{ d(z^a,z^n)\},label\{z^a\} \neq label\{z^p\}$. $z^n$ is defined as the closest vector that from a different cluster. The objective immediately follows,

\begin{equation}\label{3}
   d(z^a,z^p)+\alpha<d(z^a,z^n), \forall(z^a,z^p,z^n) \in\mathbb{Z}.
\end{equation}

Above, $\alpha$ is the margin defined accordingly by us. Thus, by the contrastive loss, we tend to optimize,

\begin{align}
    L_{con} &= \mathbb{E}_{z^a \sim Z}[d(z^a,z^p)-d(z^a,z^n)+\alpha]. \nonumber \\
    L_{con} &= \mathbb{E}_{x \sim \{RNA\}}[d(E_{j}(x),z^p)-d(E_{j}(x),z^n)+\alpha] \nonumber \\
    +&{E}_{x \sim \{ATAC\}}[d(E_{i}(x),z^p)-d(E_{i}(x),z^n)+\alpha]. 
\end{align}\label{4}

Figure \ref{fig:cycle&contrastive}b shows after training, instances within the same batch are pushed towards each other, and those from the different batches are pushed away. Thus, the decision boundary of the labels tends to be smoother and robust, which also benefits the alignment task.

\subsection{Training Procedure}
In the above sections, we proposed several losses related to different objectives. Following the training procedure of Generative Adversarial Nets \cite{goodfellow2014generative}, we adopt a two-stage training scheme where $L_{adv}$ and $L_{recon},L_{cyc},L_{con}$ are trained separately as the pseudo-code in Algorithm \ref{alg:cap}.

\begin{algorithm}
\caption{Training Procedure}\label{alg:cap}
\begin{algorithmic}

\WHILE{numbers of training iterations}
\WHILE{$k_1$ steps}
    \STATE sample mini-batch $\{x_{1},x_{2},...,x_{m}\}$ from \{scRNA-seq\}
    \STATE sample mini-batch $\{y_{1},y_{2},...,y_{m}\}$ from \{scATAC-seq\}
    \STATE Search positives and negatives $z^a,z^p$ for each $x_{1},...,y_{m}$.
    \STATE Update $E_{i},E_{j},D_{i},D_{j}$ by descending its stochastic gradient $\frac{1}{m}\nabla L_{recon}+L_{cyc}+L_{con}-L_{adv}$
\ENDWHILE
\WHILE{$k_2$ steps}
    \STATE sample mini-batch $\{x_{1},x_{2},...,x_{m}\}$ from \{scRNA-seq\}
    \STATE sample mini-batch $\{y_{1},y_{2},...,y_{m}\}$ from \{scATAC-seq\}
    \STATE Update Discriminator $\mathscr{D}$ by descending its stochastic gradient $\frac{1}{2m}\nabla L_{adv}$
\ENDWHILE
\ENDWHILE
\end{algorithmic}
\end{algorithm}

In this way, the Discriminator $\mathscr{D}$ competes against the encoder-decoder $E_{i},E_{j},D_{i},D_{j}$ until the training ends and reaches the equilibrium.

\section{Experimental Setup}

\subsubsection{Real-world Dataset.}We use two sets of single-cell multi-omics data generated by co-assays. The first dataset is generated using the sci-CAR assay \cite{cao2018joint}. For the single-cell ATAC-seq data, we download the processed data from \cite{dai2021multi}, which was computed as described in \cite{cao2018joint}, resulting in a matrix of 1791$\times$815. For the single-cell RNA-seq data, we pick the genes with $\mathnormal{q}-value>0.05$ from the genes being differentially expressed \cite{cao2018joint}, which forms a 1791$\times$2613 matrix. Such paried data was collected from human lung adenocarcinoma-derived A549 cells corresponding to 0-, 1-, or 3-hour treatment with DEX, so we have three batches here and the batch label information is available.

We denote the results of SNAREseq \cite{chen2019high} assay as the second datasets, which also consists of chromatin accessibility and gene expression. The data was collected from a mixture of human cell lines: BJ, H1, K562, and GM12878. We reduce the dimension of the data by PCA. The resulting matrix for ATAC is of size $1047\times1000$ and $1047\times500$ for gene matrix. Annotation information for BJ, H1, K562, and GM12878 is generated by the code provided by author, so labels of four batches is available.

\subsubsection{Simulated Datasets.} We simulate several datasets of different sizes, which contain 1200, 2100, 3000, and 6000 cells, respectively. For the single-cell RNA-seq data, we utilize three Gaussian distributions with different parameters to generate three batches, and the feature dimension is 1000. For the single-cell ATAC-seq data, we train an autoencoder with the simulated RNA-seq data and map them to 500 dimensions as ATAC-seq data. After that, we randomly set around 40\% of features to 0 for RNA-seq data since the real-world RNA-seq data matrix is very sparse. Considering the inevitable mismatch in the experiment, we also randomly set around 10\% mismatches in the datasets and shuffle all the pairs. Furthermore, we add noise to them with the SNR equal to 5, 10, 15, 20, and 25, along with the version without noise. Then, we have 24 simulated datasets here, and their parameters are close to those of the real-world dataset. Surely, the real-world multi-omics are more complicated than the simulated data, but the experimental results show that our method is sufficient to distinguish the performance of different methods.

\subsubsection{Evaluation criteria.} 
We utilize two existing manners \cite{dai2021multi} to evaluate integration and alignment, respectively. (a) The fraction of cells whose batch assignment is predicted correctly based on the latent space embedding. (b) $\mathnormal{recall@k}$, \textit{i.e.}, the proportion of cells whose true match is within the $\mathnormal{k}$ closest samples in the embedding space (in $\ell_1$-distance) or in the original space for methods that do not generate an embedding space. (a) is for integration, while (b) is for alignment.
\section{Results}
  \begin{figure}[!t]
    \centering  
    \subfigure[Integration performance on sci-CAR]{
    
    \includegraphics[width=0.45\textwidth]{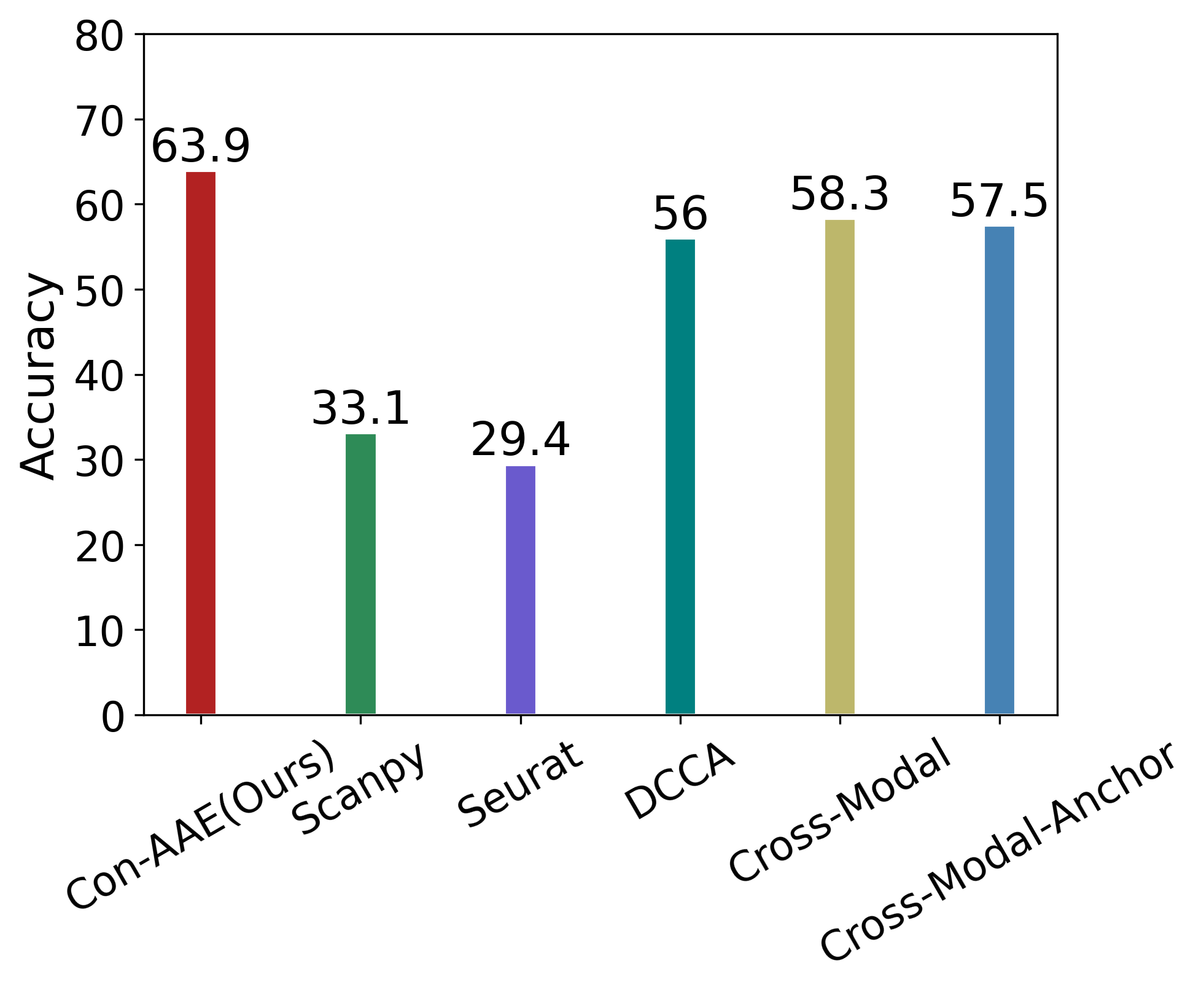}}
    \subfigure[Alignment performance on sci-CAR]{
    
    \includegraphics[width=0.45\textwidth]{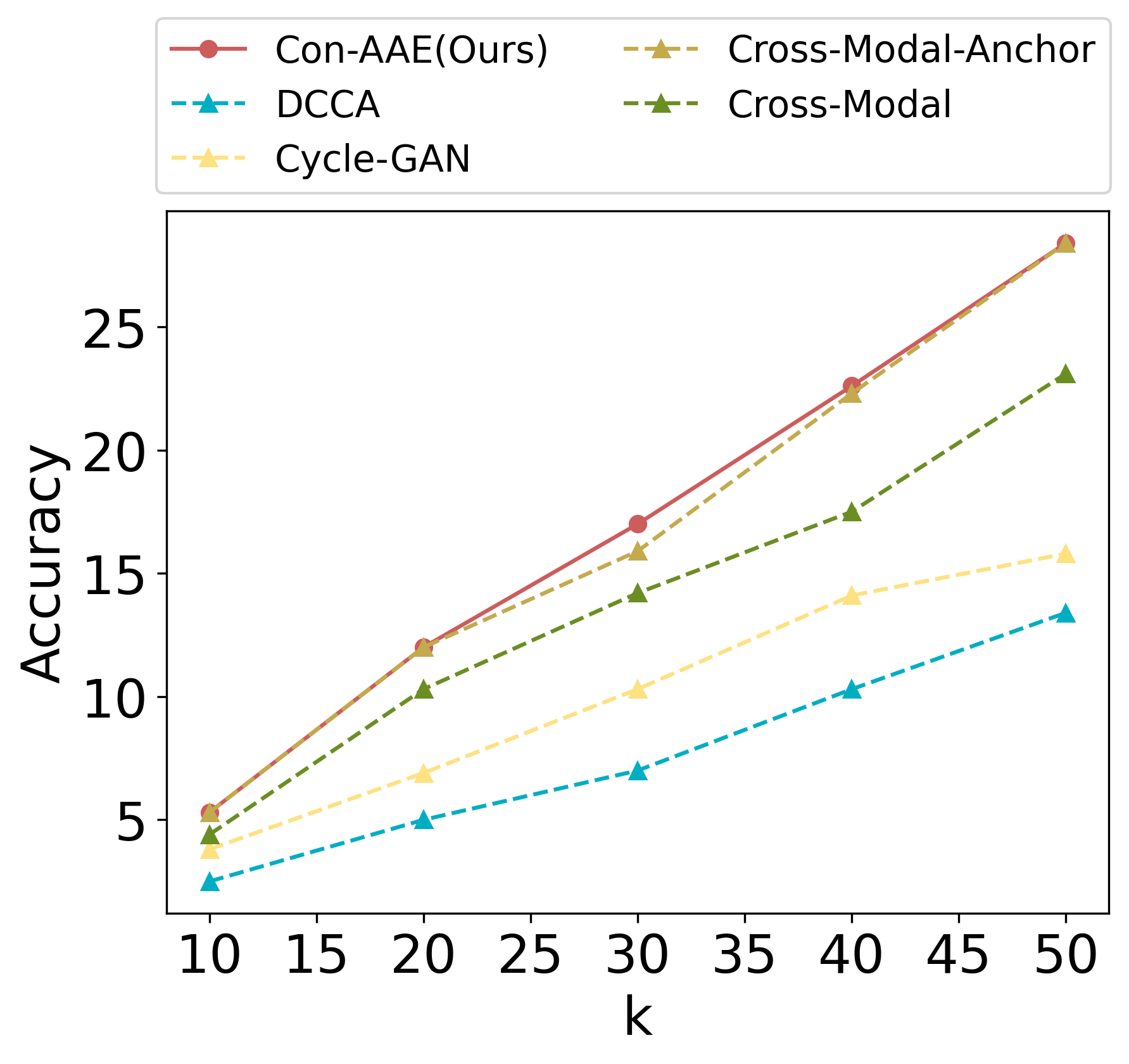}}
    
    \subfigure[Integration performance on SNAREseq]{
    \includegraphics[width=0.45\textwidth]{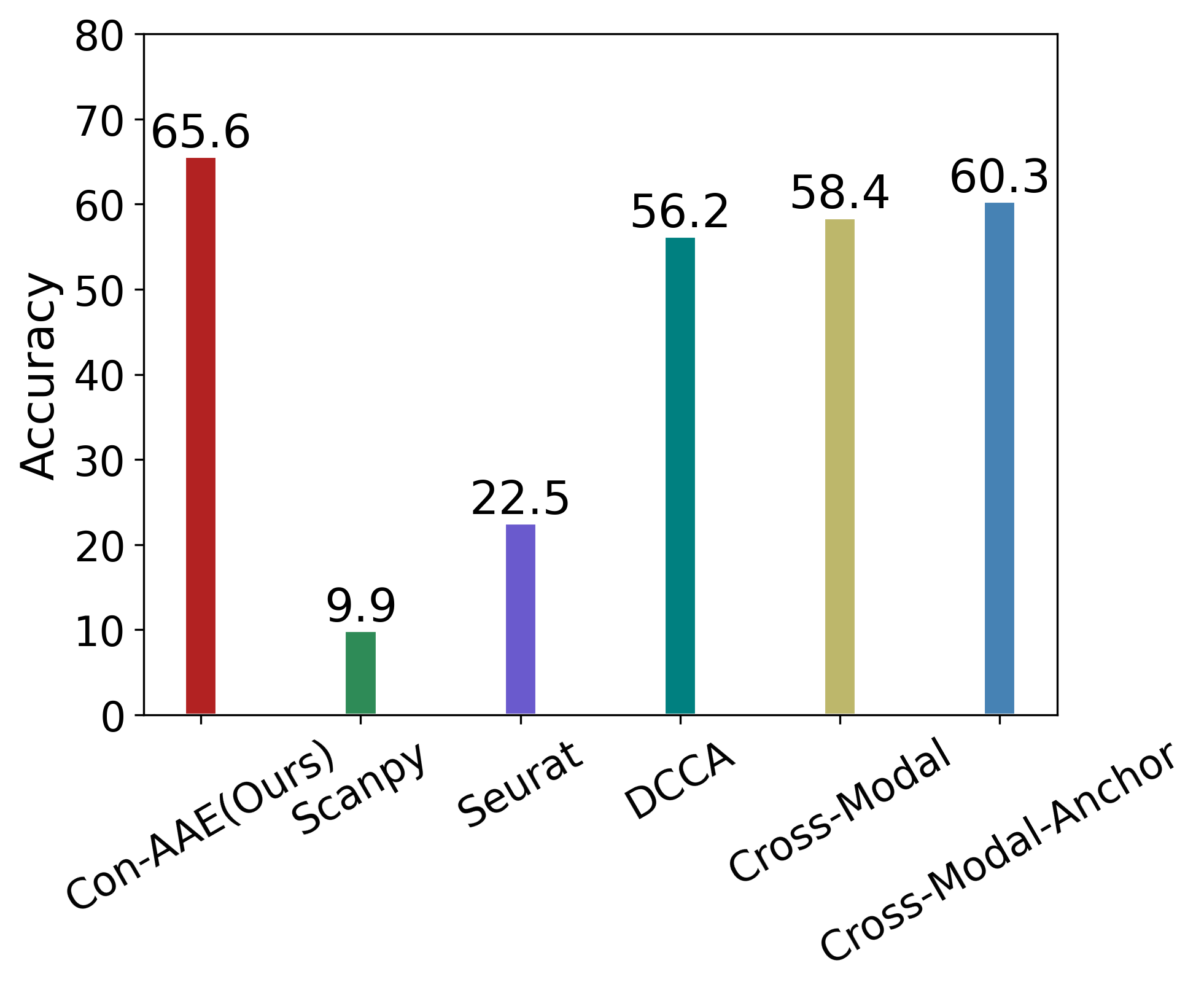}}
    \subfigure[Alignment performance on SNAREseq]{
    
    \includegraphics[width=0.45\textwidth]{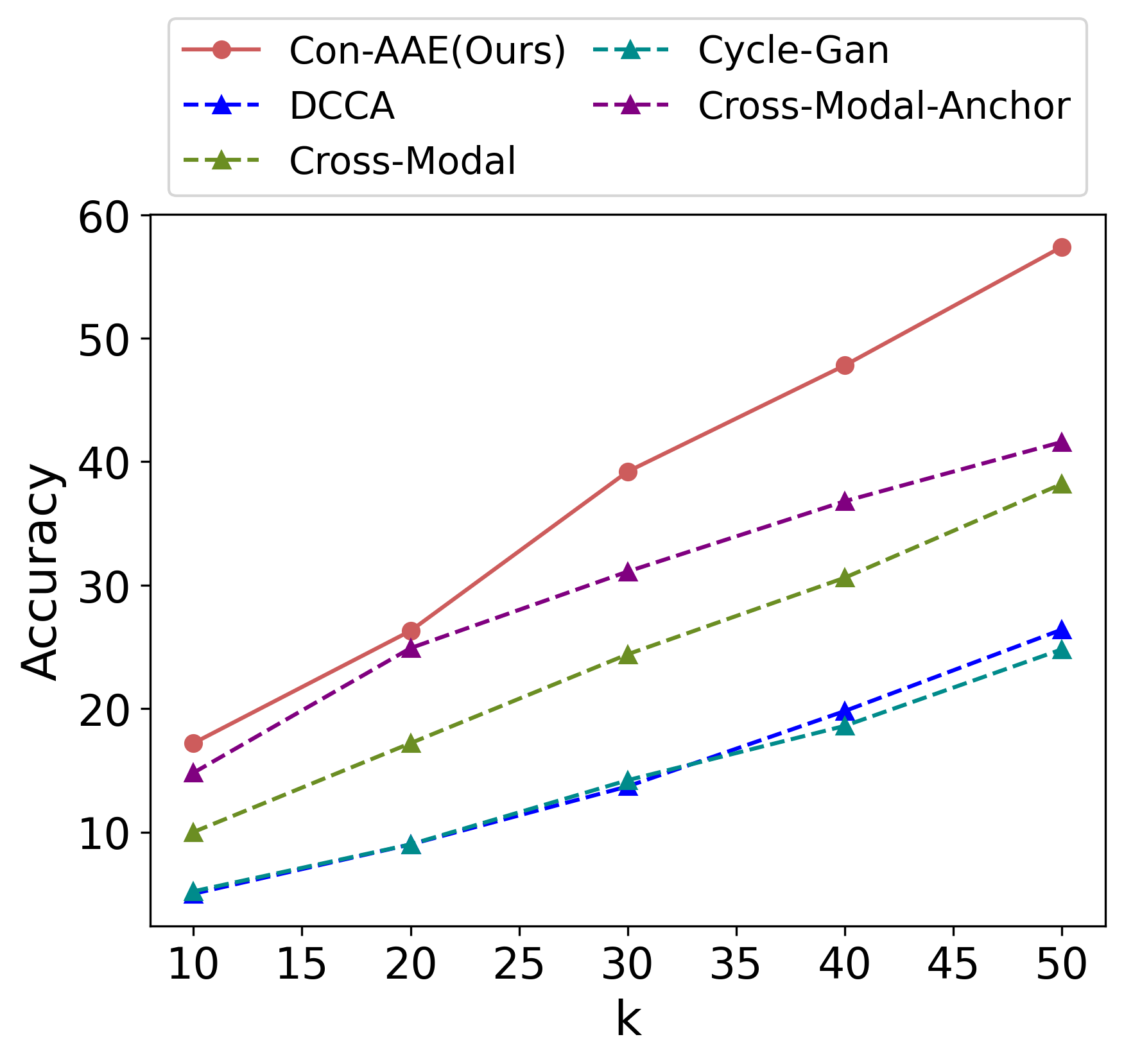}}

    \caption{Con-AAE compared with SOTA methods on the real-world datasets. Con-AAE has the best performance on both criteria. Note that the identification of cell pairwise correspondences between single cells is termed ``anchor''\cite{stuart2019comprehensive}. Cross-modal-anchor indicates that ``anchor'' information is provided when training Cross-modal.}
    \label{fig:real-results}
\end{figure}

\begin{figure*}[!t]
    \centering
    \includegraphics[width=1\textwidth]{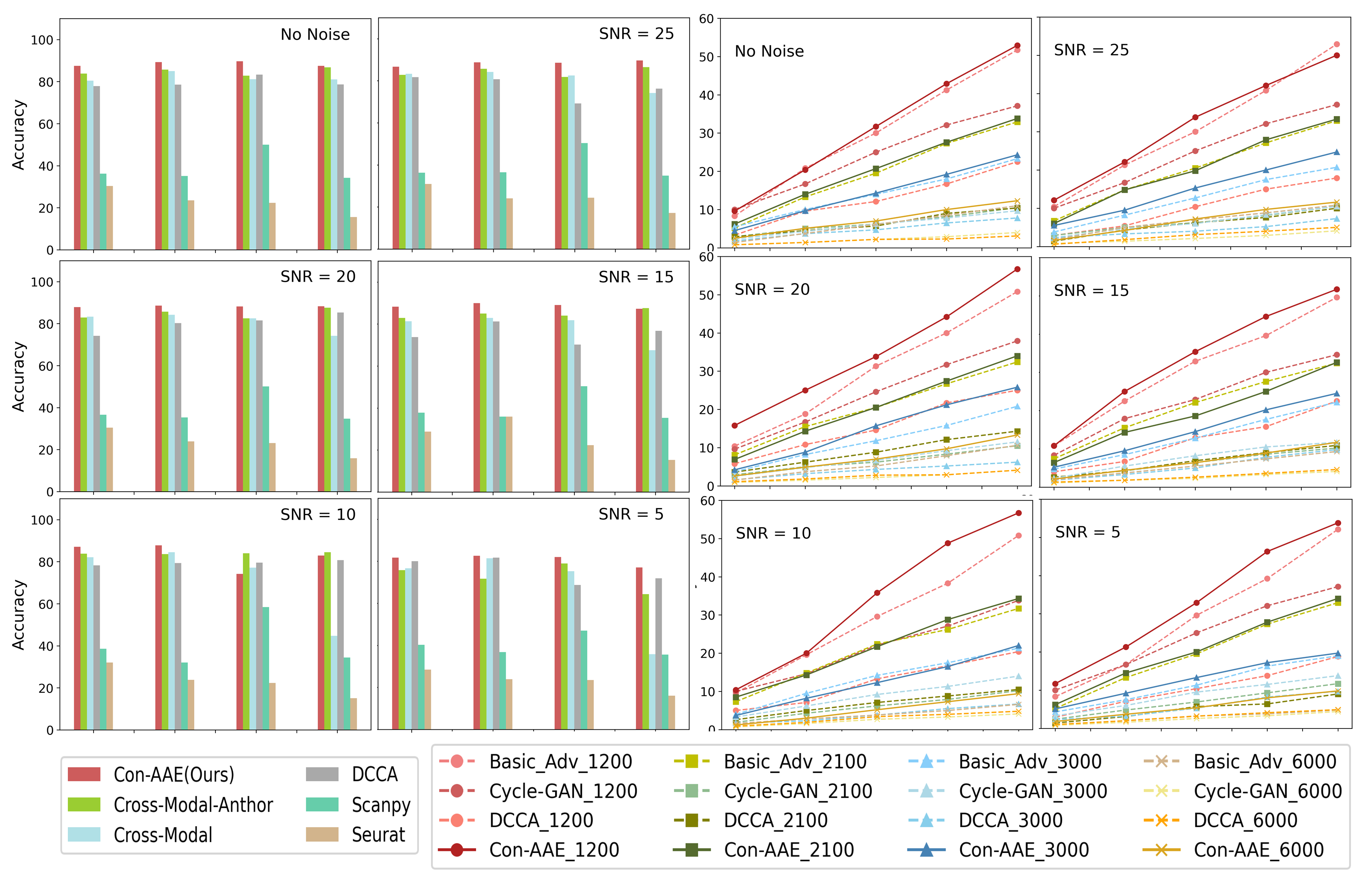}
    \centering
    \caption{Con-AAE compared with SOTA methods on simulated datasets. Con-AAE almost always has the best performance with the change of data size and SNR, while the other methods struggle with SNR or data size changing. Because there is no embedding space in Cycle-GAN, we are unable to evaluate it on integration. cross-modal-anchor is a supervised method (pairwise information provided), so we do not compare with it on alignment.}
    \label{fig:simu}
\end{figure*}

\begin{table*}[htbp]
\renewcommand\arraystretch{0.5}
\centering
\caption{\label{tab:abla-real}Ablation Study of different components in Con-AAE and Comparison with other methods. Basic refers to Coupled AEs plus Simple classifier; Adv refers to adversarial loss; mmd refers to mmd loss; anchor refers to pairwise information added; cyc refers to latent cycle-consistency loss; contra refers to contrastive loss.
}


	\begin{spacing}{1.1}
	\end{spacing}{}
	\begin{spacing}{1.15}

	\begin{tabular}{ccccccc}
	\toprule[2pt]

	\multirow{2}*{Method}	 & 
	Integration  &	Recall@k & Recall@k & Recall@k & Recall@k & Recall@k \\
	 & ACC & k=10 & k=20 & k=30 & k=40 & k=50\\
	\hline
	
	Basic &	56.1 &	3.9 & 	9.2	& 13.4	& 18.1	& 20.6\\
	Basic\_anchor &	60.8 &	5.3	& 12 &	15.9 &	22.3 &	28.4\\
	Basic\_cyc &	56.7 &	6.7&	11.7&	15.6&	19.8	&24.5\\
	Basic\_contra &	58.9 &	6.9&	12.5 &	15.6 &	21.7 &	27.3\\
	Basic\_contra\_cyc & 62.2 &	7.8&	10.8&	15.9	&21.2	&25.9\\
	\hline

	Basic\_mmd &	57.5 &	4.7 &	9.4	&12.5	&17.3 & 21.2\\
	Basic\_mmd\_anchor&	56.4	&5.3	&11.4&	16.7&	21.2&	24.3\\
	Basic\_mmd\_cyc&	60.8&	3.9&	10&	16.2&	21.2&	25.1\\
	Basic\_mmd\_contra&	57.5&	5.8&	12.8&	15.6&	24&	27\\
	Basic\_mmd\_contra\_cyc & 57.8 &	4.5&	9.5&	13.7	&19.8	&24.8\\
		
	\hline
	
	Basic\_adv &	58.3 &	4.4&	10.3&	14.2&	17.5&	23.1\\
	Basic\_adv\_anchor&	61.7&	5	&10.3&	15.9&	21.2&	26.5\\
	Basic\_adv\_cyc&	58.7&	5&	10.3&	14.2&	19.8&	24.3\\
	Basic\_adv\_contra&	60.61&	4.4	&9.4&	13.6&	18.4&	25.4\\
	\textbf{Con-AAE} & \textbf{63.9} &	\textbf{5.3}&	\textbf{12}&	\textbf{17}	&\textbf{22.6}	&\textbf{28.4}\\

	\bottomrule[2pt]
	\end{tabular}
	\end{spacing}

\end{table*}

\subsection{Compared with SOTA.}
Instead of assuming all datasets share the same underlying structure or specifying parts of hyperparameters like some traditional machine learning mathods \cite{demetci2020gromov, welch2017matcher,cao2020unsupervised,singh2020unsupervised,cao2020manifold,stuart2019comprehensive}, we obtain more information from datasets with partial correspondence information (batch label or cell types label).  We selected several state-of-art methods based on deep learning like us, including cross-modal \cite{dai2021multi}, cross-modal-anchor (pairwise information added),  DCCA \cite{andrew2013deep}, cycle-GAN \cite{zhu2017unpaired}. Moreover, we also compared our method with two classic machine learning methods of integration, which are Scanpy\cite{wolf2018scanpy} and Seurat\cite{stuart2019comprehensive}. These two methods are under the assumption that all datasets have the same features, so we applied PCA to make datasets have same features before using them. We applied Con-AAE and these methods on the simulated datasets and the real-world dataset, respectively.  

\subsubsection{On Simulated Datasets.} 
The extensive experimental results are shown in Figure \ref{fig:simu}. As shown in left part of Figure \ref{fig:simu}, Con-AAE performs better than all the other methods in most cases on the integration task, regardless of data size and SNR. Regarding alignment, we evaluate different methods using $\mathnormal{recall@k}$, whose results are shown in right part of Figure \ref{fig:simu}. Again, Con-AAE is consistently better than the other competing methods. 
Notice that our method's performance is very consistent against different data sizes and noise levels, while the other methods may perform well on some settings but badly for the other settings. 
The results indicate that Con-AAE is robust and stable enough to have the potential to handle the complicated single-cell multi-omics alignment and integration problems with a low SNR ratio.

\subsubsection{On Real-world Dataset.}
Eventually, we care about the methods' performance on the real-world dataset the most, although the real-world dataset with ground-truth information is very limited. Still, Con-AAE shows superior performance. 
On the sci-CAR datasets. Con-AAE outperforms the other methods by up to 34.5$\%$ on the integration task, as shown in Figure \ref{fig:real-results}a. For alignment, Con-AAE always has better performance than all the other methods no matter what k is (Figure \ref{fig:real-results}c). On the SNAREseq datasets, more obviously, Con-AAE also has dominant performance on each evaluation metric. The improvement on the integration task is up to 55.7$\%$ (Figure \ref{fig:real-results}b). On the other hand, the performance on $\mathnormal{recall@k}$ is better than others no matter what k is (Figure \ref{fig:real-results}d).

\subsection{Ablation Studies}
We perform comprehensive ablation studies on the Sci-CAR dataset, and the results shows the effectiveness of different components. 



There are three parts in Table \ref{tab:abla-real}. The first part indicates there is no adversarial loss in embedding space. The second part indicates an MMD loss \cite{binkowski2018demystifying} instead of an adversarial loss. And the last part indicates whether there is an adversarial loss in the embedding space. Clearly, all the items in the third part are better than the corresponding items of the other two parts, demonstrating that the adversarial loss works better than MMD loss on this problem. 

There are five items in each part of Table \ref{tab:abla-real}. The first one is the basic framework, as mentioned before. The anchor one means pairwise information provided, which indicates that it is a supervised learning problem instead of an unsupervised one. 
cyc and contra denote latent cycle-consistency loss and contrastive loss, respectively. As shown in the table, the addition of cyc and contra improves the model to some extent. Apparently, Con-AAE has the best performance. Latent cycle-consistency loss and contrastive loss alone can improve the performance to some degree, but Con-AEE is more robust and has better scalability. 

Impressively, Con-AAE has better performance even compared to some supervised methods with the pairwise information provided. Within Table \ref{tab:abla-real}, we compare our method with the anchor methods. For such methods, we train the basic coupled autoencoders using the pairwise information as the supervision. For Con-AAE, we still perform unsupervised learning using cycle-consistency loss and contrastive loss. Even without the supervised information, Con-AAE can still outperform the basic supervised anchor methods consistently on both accuracy and $\mathnormal{recall@k}$. It suggests that cycle-consistency loss and contrastive loss can force our method to learn a unified latent space for the two kinds of single-cell omics data, making the alignment and integration much easier. We also tried to combine Con-AAE with the pairwise information. The supervised information can help our method further, but the degree is very slight. We suppose that in the real data, the pairwise information may contain noise, which is common in the single-cell field. Because of the contrastive loss, which makes Con-AAE a robust method, such weak supervision does not help our method too much.

\begin{figure}[!t]
    \begin{center}
    \includegraphics[width=0.9\textwidth]{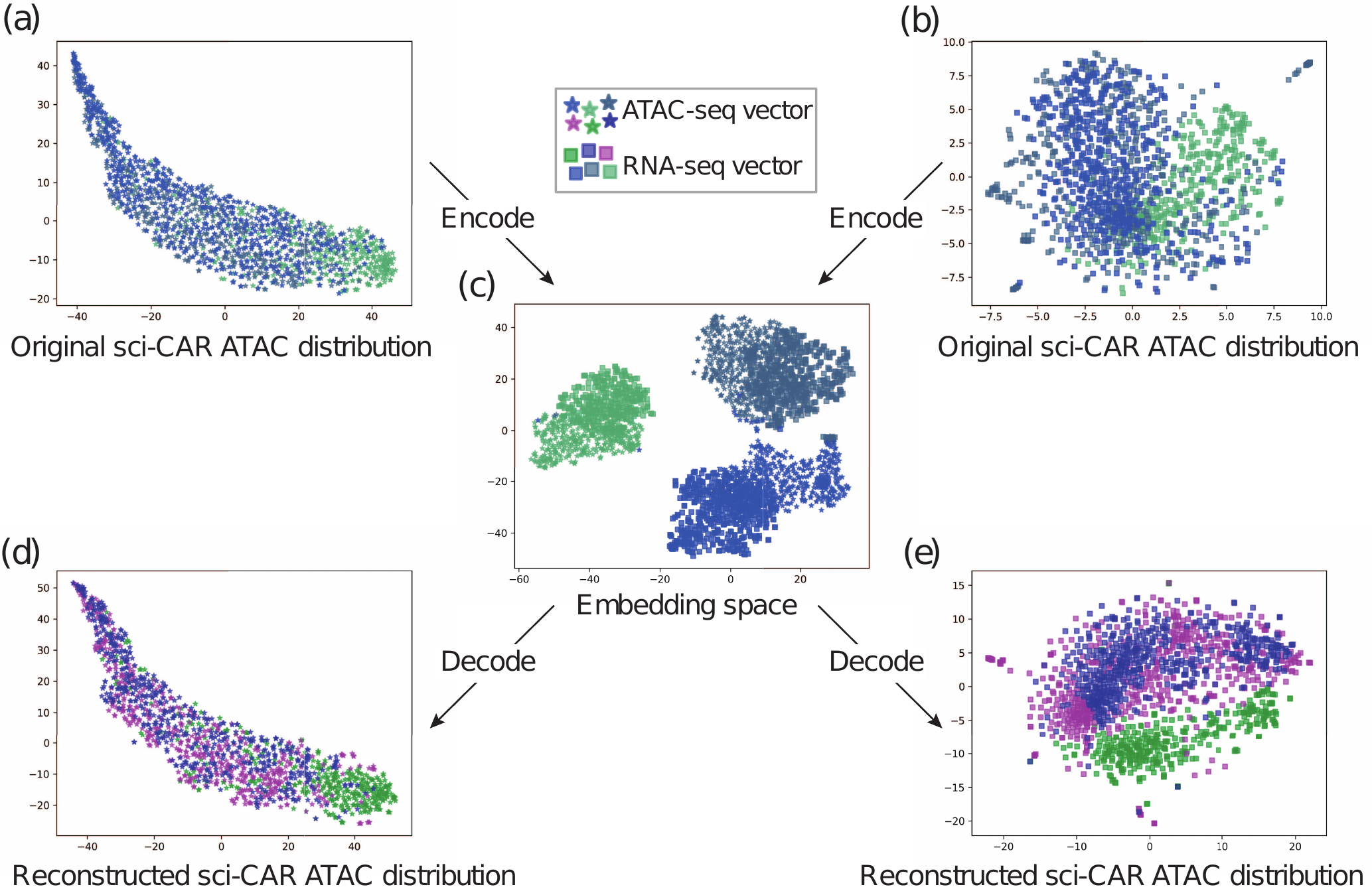}
    \end{center}
    \caption{Visualization of the sci-CAR data in the original spaces, shared embedding space, and the reconstructed spaces. The star represents the scATAC-seq data while the square means the scRNA-seq data. Different colors refer to different batch labels, corresponding to 0-, 1-, 3- hours treatment with DEX. For easy viewing, we used different colors when reconstructing data. The distribution of reconstructed data is still distinguishable.}
    \label{fig:sci_CAR_Visual}
\end{figure}

\subsection{Visualization}
\begin{figure}[!t]
    \begin{center}
    \includegraphics[width=0.9\textwidth]{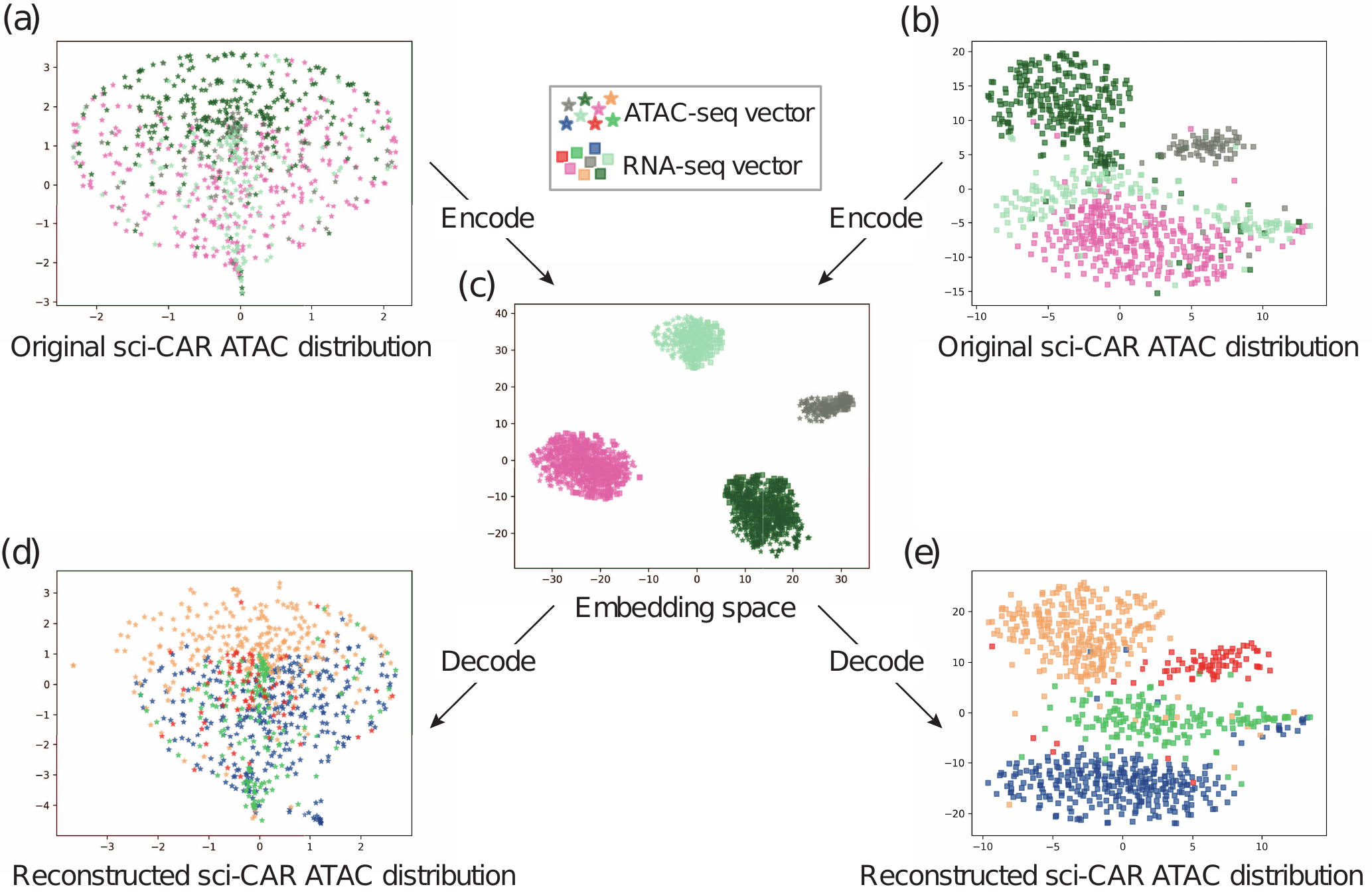}
    \end{center}
    \caption{Visualization of the SNAREseq data in the original spaces, shared embedding space, and the reconstructed spaces. The star represents the scATAC-seq data while square means the scRNA-seq data. Different colors refer to the H1, BJ, K562, and GM12878 human cells. We can see that the scATAC-seq data and scRNA-seq data from the same cell type are tightly combined in the embedding space, And also, we utilize different colors representing reconstructed data, which also has an impressive effect.}
    \label{fig:SNAREseq_Visual}
\end{figure}

To visualize the integration and alignment, we utilize t-SNE to project the data from the embedding space to the 2D space. As we can see in Figure \ref{fig:sci_CAR_Visual} and \ref{fig:SNAREseq_Visual}, for the two real-world datasets, we projected the scATAC-seq data and scRNA-seq data to a shared embedding space. And within the space, the different modality (with different shapes) data from the same batch (with the same color) form into clusters, suggesting that our method indeed learned a latent space where the data from different omics can be integrated easily. Furthermore, we used the decoder of each side to translate the embedding vector back to the original space. The resulted data distribution is quite close to the original distribution, which indicates that our framework indeed learned the underlying features of the data and removed the redundant features effectively in the process of encoding.

\section{Discussion}
  In this paper, we propose a novel framework, Con-AAE, aiming at integrating and aligning the multi-omics data at the single-cell level. On the one hand, our proposed method can map different modalities into the embedding spaces and overlap these two distributions with the help of an adversarial loss and a novel latent cycle-consistency loss. 
On the other hand, we apply a novel self-supervised contrastive loss in the embedding space to improve the robustness and scalability of the entire framework. Comprehensive experimental results on both the simulated datasets and the real datasets show that the proposed framework can outperform the other state-of-the-art methods for both the alignment and the integration tasks. 
Detailed ablation studies also dissect and demonstrate the effectiveness of each component in the framework.
Our method will be helpful for both the single-cell multi-omics research and the general multi-modality learning tasks in computational biology.

For future work, we aim to extend our work from a two-domain task to a multiple-domain study, allowing it to integrate and align multiple modalities. Besides integration and alignment between sequence modalities, we intend to perform our method on different kinds of data, including but not limited to images, geometrical spatial structure, \textit{etc}. Obviously, it is very interesting to investigate the spatial transcriptomics data. We will also develop methods for translating modalities. By doing so, we hope to build a system that could be utilized in medical, biological, and other fields.

\section{Acknowledgements} 
This work was supported by a grant by The CUHK Shenzhen Research Institute. Thanks to Yixuan Wang for providing partial exquisite images.

\bibliographystyle{myrecomb}

\bibliography{main}

\begin{thebibliography}{10}
\expandafter\ifx\csname url\endcsname\relax
  \def\url#1{\texttt{#1}}\fi
\expandafter\ifx\csname urlprefix\endcsname\relax\def\urlprefix{URL }\fi
\providecommand{\bibinfo}[2]{#2}
\providecommand{\eprint}[2][]{\url{#2}}

\bibitem{baltruvsaitis2018multimodal}
\bibinfo{author}{Baltru{\v{s}}aitis, T.}, \bibinfo{author}{Ahuja, C.} \&
  \bibinfo{author}{Morency, L.-P.}
\newblock \bibinfo{title}{Multimodal machine learning: A survey and taxonomy}.
\newblock \emph{\bibinfo{journal}{IEEE transactions on pattern analysis and
  machine intelligence}} \textbf{\bibinfo{volume}{41}},
  \bibinfo{pages}{423--443} (\bibinfo{year}{2018}).

\bibitem{gala2021consistent}
\bibinfo{author}{Gala, R.} \emph{et~al.}
\newblock \bibinfo{title}{Consistent cross-modal identification of cortical
  neurons with coupled autoencoders}.
\newblock \emph{\bibinfo{journal}{Nature Computational Science}}
  \textbf{\bibinfo{volume}{1}}, \bibinfo{pages}{120--127}
  (\bibinfo{year}{2021}).

\bibitem{cao2018joint}
\bibinfo{author}{Cao, J.} \emph{et~al.}
\newblock \bibinfo{title}{Joint profiling of chromatin accessibility and gene
  expression in thousands of single cells}.
\newblock \emph{\bibinfo{journal}{Science}} \textbf{\bibinfo{volume}{361}},
  \bibinfo{pages}{1380--1385} (\bibinfo{year}{2018}).

\bibitem{klein2015droplet}
\bibinfo{author}{Klein, A.~M.} \emph{et~al.}
\newblock \bibinfo{title}{Droplet barcoding for single-cell transcriptomics
  applied to embryonic stem cells}.
\newblock \emph{\bibinfo{journal}{Cell}} \textbf{\bibinfo{volume}{161}},
  \bibinfo{pages}{1187--1201} (\bibinfo{year}{2015}).

\bibitem{macosko2015highly}
\bibinfo{author}{Macosko, E.~Z.} \emph{et~al.}
\newblock \bibinfo{title}{Highly parallel genome-wide expression profiling of
  individual cells using nanoliter droplets}.
\newblock \emph{\bibinfo{journal}{Cell}} \textbf{\bibinfo{volume}{161}},
  \bibinfo{pages}{1202--1214} (\bibinfo{year}{2015}).

\bibitem{buenrostro2015single}
\bibinfo{author}{Buenrostro, J.~D.} \emph{et~al.}
\newblock \bibinfo{title}{Single-cell chromatin accessibility reveals
  principles of regulatory variation}.
\newblock \emph{\bibinfo{journal}{Nature}} \textbf{\bibinfo{volume}{523}},
  \bibinfo{pages}{486--490} (\bibinfo{year}{2015}).

\bibitem{butler2018integrating}
\bibinfo{author}{Butler, A.}, \bibinfo{author}{Hoffman, P.},
  \bibinfo{author}{Smibert, P.}, \bibinfo{author}{Papalexi, E.} \&
  \bibinfo{author}{Satija, R.}
\newblock \bibinfo{title}{Integrating single-cell transcriptomic data across
  different conditions, technologies, and species}.
\newblock \emph{\bibinfo{journal}{Nature biotechnology}}
  \textbf{\bibinfo{volume}{36}}, \bibinfo{pages}{411--420}
  (\bibinfo{year}{2018}).

\bibitem{trong2020semisupervised}
\bibinfo{author}{Trong, T.~N.} \emph{et~al.}
\newblock \bibinfo{title}{Semisupervised generative autoencoder for single-cell
  data}.
\newblock \emph{\bibinfo{journal}{Journal of Computational Biology}}
  \textbf{\bibinfo{volume}{27}}, \bibinfo{pages}{1190--1203}
  (\bibinfo{year}{2020}).

\bibitem{stuart2019comprehensive}
\bibinfo{author}{Stuart, T.} \emph{et~al.}
\newblock \bibinfo{title}{Comprehensive integration of single-cell data}.
\newblock \emph{\bibinfo{journal}{Cell}} \textbf{\bibinfo{volume}{177}},
  \bibinfo{pages}{1888--1902} (\bibinfo{year}{2019}).

\bibitem{bersanelli2016methods}
\bibinfo{author}{Bersanelli, M.} \emph{et~al.}
\newblock \bibinfo{title}{Methods for the integration of multi-omics data:
  mathematical aspects}.
\newblock \emph{\bibinfo{journal}{BMC bioinformatics}}
  \textbf{\bibinfo{volume}{17}}, \bibinfo{pages}{167--177}
  (\bibinfo{year}{2016}).

\bibitem{argelaguet2018multi}
\bibinfo{author}{Argelaguet, R.} \emph{et~al.}
\newblock \bibinfo{title}{Multi-omics factor analysis—a framework for
  unsupervised integration of multi-omics data sets}.
\newblock \emph{\bibinfo{journal}{Molecular systems biology}}
  \textbf{\bibinfo{volume}{14}}, \bibinfo{pages}{e8124} (\bibinfo{year}{2018}).

\bibitem{stanley2020harmonic}
\bibinfo{author}{Stanley~III, J.~S.}, \bibinfo{author}{Gigante, S.},
  \bibinfo{author}{Wolf, G.} \& \bibinfo{author}{Krishnaswamy, S.}
\newblock \bibinfo{title}{Harmonic alignment}.
\newblock In \emph{\bibinfo{booktitle}{Proceedings of the 2020 SIAM
  International Conference on Data Mining}}, \bibinfo{pages}{316--324}
  (\bibinfo{organization}{SIAM}, \bibinfo{year}{2020}).

\bibitem{andrew2013deep}
\bibinfo{author}{Andrew, G.}, \bibinfo{author}{Arora, R.},
  \bibinfo{author}{Bilmes, J.} \& \bibinfo{author}{Livescu, K.}
\newblock \bibinfo{title}{Deep canonical correlation analysis}.
\newblock In \emph{\bibinfo{booktitle}{International conference on machine
  learning}}, \bibinfo{pages}{1247--1255} (\bibinfo{organization}{PMLR},
  \bibinfo{year}{2013}).

\bibitem{cao2020manifold}
\bibinfo{author}{Cao, K.}, \bibinfo{author}{Hong, Y.} \& \bibinfo{author}{Wan,
  L.}
\newblock \bibinfo{title}{Manifold alignment for heterogeneous single-cell
  multi-omics data integration using pamona}.
\newblock \emph{\bibinfo{journal}{bioRxiv}}  (\bibinfo{year}{2020}).

\bibitem{borgwardt2006integrating}
\bibinfo{author}{Borgwardt, K.~M.} \emph{et~al.}
\newblock \bibinfo{title}{Integrating structured biological data by kernel
  maximum mean discrepancy}.
\newblock \emph{\bibinfo{journal}{Bioinformatics}}
  \textbf{\bibinfo{volume}{22}}, \bibinfo{pages}{e49--e57}
  (\bibinfo{year}{2006}).

\bibitem{welch2017matcher}
\bibinfo{author}{Welch, J.~D.}, \bibinfo{author}{Hartemink, A.~J.} \&
  \bibinfo{author}{Prins, J.~F.}
\newblock \bibinfo{title}{Matcher: manifold alignment reveals correspondence
  between single cell transcriptome and epigenome dynamics}.
\newblock \emph{\bibinfo{journal}{Genome biology}}
  \textbf{\bibinfo{volume}{18}}, \bibinfo{pages}{1--19} (\bibinfo{year}{2017}).

\bibitem{singh2020unsupervised}
\bibinfo{author}{Singh, R.} \emph{et~al.}
\newblock \bibinfo{title}{Unsupervised manifold alignment for single-cell
  multi-omics data}.
\newblock In \emph{\bibinfo{booktitle}{Proceedings of the 11th ACM
  International Conference on Bioinformatics, Computational Biology and Health
  Informatics}}, \bibinfo{pages}{1--10} (\bibinfo{year}{2020}).

\bibitem{cao2020unsupervised}
\bibinfo{author}{Cao, K.}, \bibinfo{author}{Bai, X.}, \bibinfo{author}{Hong,
  Y.} \& \bibinfo{author}{Wan, L.}
\newblock \bibinfo{title}{Unsupervised topological alignment for single-cell
  multi-omics integration}.
\newblock \emph{\bibinfo{journal}{Bioinformatics}}
  \textbf{\bibinfo{volume}{36}}, \bibinfo{pages}{i48--i56}
  (\bibinfo{year}{2020}).

\bibitem{demetci2020gromov}
\bibinfo{author}{Demetci, P.}, \bibinfo{author}{Santorella, R.},
  \bibinfo{author}{Sandstede, B.}, \bibinfo{author}{Noble, W.~S.} \&
  \bibinfo{author}{Singh, R.}
\newblock \bibinfo{title}{Gromov-wasserstein optimal transport to align
  single-cell multi-omics data}.
\newblock \emph{\bibinfo{journal}{BioRxiv}}  (\bibinfo{year}{2020}).

\bibitem{li2019deep}
\bibinfo{author}{Li, Y.} \emph{et~al.}
\newblock \bibinfo{title}{Deep learning in bioinformatics: Introduction,
  application, and perspective in the big data era}.
\newblock \emph{\bibinfo{journal}{Methods}} \textbf{\bibinfo{volume}{166}},
  \bibinfo{pages}{4--21} (\bibinfo{year}{2019}).

\bibitem{li2020modern}
\bibinfo{author}{Li, H.} \emph{et~al.}
\newblock \bibinfo{title}{Modern deep learning in bioinformatics}.
\newblock \emph{\bibinfo{journal}{Journal of molecular cell biology}}
  \textbf{\bibinfo{volume}{12}}, \bibinfo{pages}{823--827}
  (\bibinfo{year}{2020}).

\bibitem{zhu2017unpaired}
\bibinfo{author}{Zhu, J.-Y.}, \bibinfo{author}{Park, T.},
  \bibinfo{author}{Isola, P.} \& \bibinfo{author}{Efros, A.~A.}
\newblock \bibinfo{title}{Unpaired image-to-image translation using
  cycle-consistent adversarial networks}.
\newblock In \emph{\bibinfo{booktitle}{Proceedings of the IEEE international
  conference on computer vision}}, \bibinfo{pages}{2223--2232}
  (\bibinfo{year}{2017}).

\bibitem{wang2017magan}
\bibinfo{author}{Wang, R.}, \bibinfo{author}{Cully, A.},
  \bibinfo{author}{Chang, H.~J.} \& \bibinfo{author}{Demiris, Y.}
\newblock \bibinfo{title}{Magan: Margin adaptation for generative adversarial
  networks}.
\newblock \emph{\bibinfo{journal}{arXiv preprint arXiv:1704.03817}}
  (\bibinfo{year}{2017}).

\bibitem{yoon2018radialgan}
\bibinfo{author}{Yoon, J.}, \bibinfo{author}{Jordon, J.} \&
  \bibinfo{author}{Schaar, M.}
\newblock \bibinfo{title}{Radialgan: Leveraging multiple datasets to improve
  target-specific predictive models using generative adversarial networks}.
\newblock In \emph{\bibinfo{booktitle}{International Conference on Machine
  Learning}}, \bibinfo{pages}{5699--5707} (\bibinfo{organization}{PMLR},
  \bibinfo{year}{2018}).

\bibitem{choi2018stargan}
\bibinfo{author}{Choi, Y.} \emph{et~al.}
\newblock \bibinfo{title}{Stargan: Unified generative adversarial networks for
  multi-domain image-to-image translation}.
\newblock In \emph{\bibinfo{booktitle}{Proceedings of the IEEE conference on
  computer vision and pattern recognition}}, \bibinfo{pages}{8789--8797}
  (\bibinfo{year}{2018}).

\bibitem{dai2021multi}
\bibinfo{author}{Dai~Yang, K.} \emph{et~al.}
\newblock \bibinfo{title}{Multi-domain translation between single-cell imaging
  and sequencing data using autoencoders}.
\newblock \emph{\bibinfo{journal}{Nature Communications}}
  \textbf{\bibinfo{volume}{12}}, \bibinfo{pages}{1--10} (\bibinfo{year}{2021}).

\bibitem{zhang2019integrated}
\bibinfo{author}{Zhang, X.} \emph{et~al.}
\newblock \bibinfo{title}{Integrated multi-omics analysis using variational
  autoencoders: Application to pan-cancer classification}.
\newblock In \emph{\bibinfo{booktitle}{2019 IEEE International Conference on
  Bioinformatics and Biomedicine (BIBM)}}, \bibinfo{pages}{765--769}
  (\bibinfo{organization}{IEEE}, \bibinfo{year}{2019}).

\bibitem{ma2019integrate}
\bibinfo{author}{Ma, T.} \& \bibinfo{author}{Zhang, A.}
\newblock \bibinfo{title}{Integrate multi-omics data with biological
  interaction networks using multi-view factorization autoencoder (mae)}.
\newblock \emph{\bibinfo{journal}{BMC genomics}} \textbf{\bibinfo{volume}{20}},
  \bibinfo{pages}{1--11} (\bibinfo{year}{2019}).

\bibitem{makhzani2015adversarial}
\bibinfo{author}{Makhzani, A.}, \bibinfo{author}{Shlens, J.},
  \bibinfo{author}{Jaitly, N.}, \bibinfo{author}{Goodfellow, I.} \&
  \bibinfo{author}{Frey, B.}
\newblock \bibinfo{title}{Adversarial autoencoders}.
\newblock \emph{\bibinfo{journal}{arXiv preprint arXiv:1511.05644}}
  (\bibinfo{year}{2015}).

\bibitem{chen2020simple}
\bibinfo{author}{Chen, T.}, \bibinfo{author}{Kornblith, S.},
  \bibinfo{author}{Norouzi, M.} \& \bibinfo{author}{Hinton, G.}
\newblock \bibinfo{title}{A simple framework for contrastive learning of visual
  representations}.
\newblock In \emph{\bibinfo{booktitle}{International conference on machine
  learning}}, \bibinfo{pages}{1597--1607} (\bibinfo{organization}{PMLR},
  \bibinfo{year}{2020}).

\bibitem{hira2021integrated}
\bibinfo{author}{Hira, M.~T.} \emph{et~al.}
\newblock \bibinfo{title}{Integrated multi-omics analysis of ovarian cancer
  using variational autoencoders}.
\newblock \emph{\bibinfo{journal}{Scientific reports}}
  \textbf{\bibinfo{volume}{11}}, \bibinfo{pages}{1--16} (\bibinfo{year}{2021}).

\bibitem{DBLP:conf/ijcai/HuW19}
\bibinfo{author}{Hu, Z.} \& \bibinfo{author}{Wang, J. T.~L.}
\newblock \bibinfo{title}{Generative adversarial networks for video prediction
  with action control}.
\newblock In \bibinfo{editor}{Seghrouchni, A. E.~F.} \& \bibinfo{editor}{Sarne,
  D.} (eds.) \emph{\bibinfo{booktitle}{Artificial Intelligence. {IJCAI} 2019
  International Workshops - Macao, China, August 10-12, 2019, Revised Selected
  Best Papers}}, vol. \bibinfo{volume}{12158} of \emph{\bibinfo{series}{Lecture
  Notes in Computer Science}}, \bibinfo{pages}{87--105}
  (\bibinfo{publisher}{Springer}, \bibinfo{year}{2019}).
\newblock \urlprefix\url{https://doi.org/10.1007/978-3-030-56150-5\_5}.

\bibitem{schroff2015facenet}
\bibinfo{author}{Schroff, F.}, \bibinfo{author}{Kalenichenko, D.} \&
  \bibinfo{author}{Philbin, J.}
\newblock \bibinfo{title}{Facenet: A unified embedding for face recognition and
  clustering}.
\newblock In \emph{\bibinfo{booktitle}{Proceedings of the IEEE conference on
  computer vision and pattern recognition}}, \bibinfo{pages}{815--823}
  (\bibinfo{year}{2015}).

\bibitem{han2021self}
\bibinfo{author}{Han, W.} \emph{et~al.}
\newblock \bibinfo{title}{Self-supervised contrastive learning for integrative
  single cell rna-seq data analysis}.
\newblock \emph{\bibinfo{journal}{bioRxiv}}  (\bibinfo{year}{2021}).

\bibitem{goodfellow2014generative}
\bibinfo{author}{Goodfellow, I.} \emph{et~al.}
\newblock \bibinfo{title}{Generative adversarial nets}.
\newblock \emph{\bibinfo{journal}{Advances in neural information processing
  systems}} \textbf{\bibinfo{volume}{27}} (\bibinfo{year}{2014}).

\bibitem{chen2019high}
\bibinfo{author}{Chen, S.}, \bibinfo{author}{Lake, B.~B.} \&
  \bibinfo{author}{Zhang, K.}
\newblock \bibinfo{title}{High-throughput sequencing of the transcriptome and
  chromatin accessibility in the same cell}.
\newblock \emph{\bibinfo{journal}{Nature biotechnology}}
  \textbf{\bibinfo{volume}{37}}, \bibinfo{pages}{1452--1457}
  (\bibinfo{year}{2019}).

\bibitem{wolf2018scanpy}
\bibinfo{author}{Wolf, F.~A.}, \bibinfo{author}{Angerer, P.} \&
  \bibinfo{author}{Theis, F.~J.}
\newblock \bibinfo{title}{Scanpy: large-scale single-cell gene expression data
  analysis}.
\newblock \emph{\bibinfo{journal}{Genome biology}}
  \textbf{\bibinfo{volume}{19}}, \bibinfo{pages}{1--5} (\bibinfo{year}{2018}).

\bibitem{binkowski2018demystifying}
\bibinfo{author}{Bi{\'n}kowski, M.}, \bibinfo{author}{Sutherland, D.~J.},
  \bibinfo{author}{Arbel, M.} \& \bibinfo{author}{Gretton, A.}
\newblock \bibinfo{title}{Demystifying mmd gans}.
\newblock \emph{\bibinfo{journal}{arXiv preprint arXiv:1801.01401}}
  (\bibinfo{year}{2018}).

\end{thebibliography}

\appendix
\section{Reproducibility}
\subsection{Details of the experiments}
The details of the network architecture are summarized in Table \ref{tab:table 2}. The learning rate of the model is 0.0001, and the batch size is 32 for the real-world dataset, 100 for simulated datasets. We trained this model for 4000 epochs using the Adam optimizer with $\beta_{1}=0.5$, $\beta_{2}=0.999$, and weight decay is set to 0.0001. Using early stopping, we found that the best model usually appears in 1000 epoch or 2000 epoch. The activation function is LeakyReLU after each layer, followed with Batch Normalization.

\begin{table}[htbp]
    \centering
    \caption{``d'' refers to the dimension of the input data. 50 is the dimension of the embedding space. Note that the last hidden layer in encoder and the fast layer in decoder are 100. We set it to be 100 because 
    the dimension is 2613 and 815 for the real-world dataset, 1000 and 500 for simulated datasets. If the dimension of the input data is larger, this number is supposed to be adjusted accordingly.}
    \label{tab:table 2}
    \begin{tabular}{|l|l|l|}
        \hline
        \multicolumn{1}{|c|}{}&Encoder&Decoder\\
        \hline
        Input size& d& 50 \\
        Hidden layer size& d, d, d, 100& 100, d, d, d \\
        Output size& 50 & d \\ 
        \hline
          
    \end{tabular}
\end{table}

The discriminator consists of 2 hidden layers with 50 and 100 nodes, respectively. The output size is 1. The simple classifier consists of 1 layer, and the ouput size is 3.
\subsection{Implementation of  Con-AAE}
We have implemented the Con-AAE in Python 3.7.7 with Pytorch 1.0, whose experiments were run on Nvidia Tesla P100. 
The source code is available at https://github.com/kakarotcq/RNA-Seq-and-ATAC-Seq-mapping. 

\subsection{The hyperparameter tunning range}
For all the methods, the weights of different loss terms are shown in Table \ref{tab:table 3}. Parts of the weights should be adjusted in accordance of the reality.  The weights of anchor loss and cycle-consistency loss are supposed to be low because the noise level of the real dataset is high, with a few mismatches, while contrastive loss should be adjusted to a  higher value, and vice versa.

\begin{table}[htbp]
    \centering
    \caption{Loss terms and their corresponding weights for training Con-AAE.}
    \label{tab:table 3}
     \begin{tabular}{|l|l|l|}
        \hline
        Loss function& Loss type & Weight\\
        \hline
        Reconstruction loss& MSELoss&10.0 \\
        Adversarial loss& MSELoss & 10.0 \\
        Classifier loss& CrossEntropyLoss & 1.0  \\
        Cycle-consistency loss& MSELoss & 1.0, 5.0, 10.0 \\
        Contrastive loss& MarginRankingLoss & 1.0, 5.0, 10.0 \\
        Anchor loss& MSELoss & 0.05, 0.1, 1.0 \\
        \hline
          
    \end{tabular}

\end{table}

\section{Extra experimental results}
 We also tried to train Con-AAE with the pairwise information, denoted as Con-AAE-anchor. The comparison between Con-AAE and Con-AAE-anchor is shown in Table \ref{tab:Con-AAEComprasion}. We can see that the performance of integration is almost the same while the performance of alignment is improved slightly as the pairwise information is added. In general, Con-AAE is a very stable unsupervised method, which is robust to noise. So the weak supervised pairwise information with a high level of noise does not improve the performance of Con-AAE very significantly. But still, this experiment suggests that Con-AAE is a flexible framework, which can incorporate supervised information. 
 
\begin{figure*}[htbp] 
\centering 
\includegraphics[width=1\textwidth]{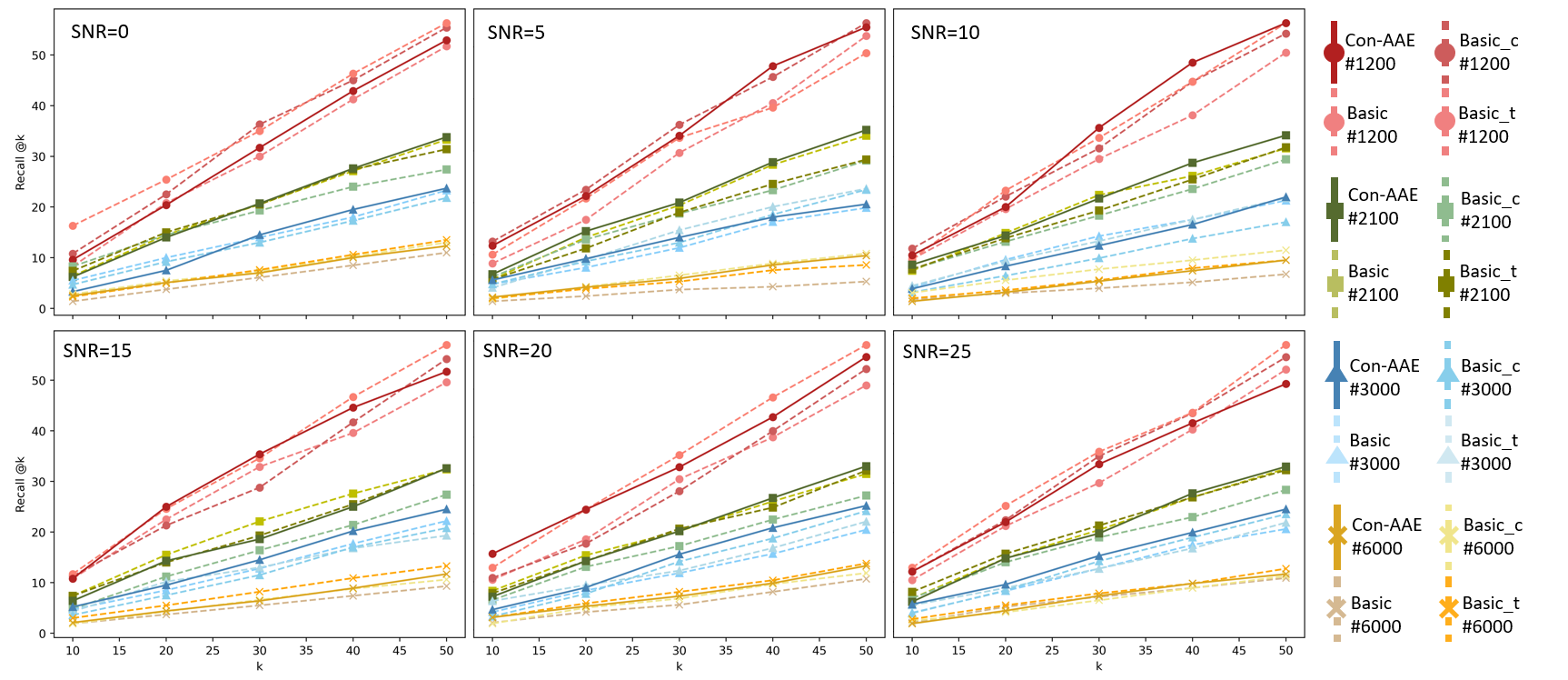} 
\caption{The alignment performance on simulated datasets with different components.} 

\label{Fig:Alignment} 
\end{figure*}

 We also conducted ablation studies on the simulated datasets. The quantitative results are shown in the Table \ref{tab:table-name} and Figure \ref{Fig:Alignment}. We can see that the cycle-consistency loss and contrastive loss could improve the model in many cases, but not stably enough. Compared with that, Con-AAE almost always has the best performance with the change of data sizes and SNRs, while the other methods struggle when the SNRs or data sizes change.
 
\begin{table*}[htbp]
    \centering
    \caption{Comparison between Con-AAE and Con-AAE-anchor on the real dataset. Con-AAE is a very stable unsupervised method, which is robust to noise. So the weak supervision with a high level of noise does not improve the performance of Con-AAE significantly. }
    \label{tab:Con-AAEComprasion}
    \begin{tabular}{|l|l|l|l|l|l|l|}
        \hline
        Method&Integration ACC& Recall@k=10&Recall@k=20&Recall@k=30&Recall@k=40&Recall@k=50\\
        \hline
        Con-AAE& 63.9&5.3&12&17&22.6&28.4 \\
        \hline
        Con-AAE-anchor& 63.4 & 7& 13.7& 18.4&25.4&29.1\\
        \hline
          
    \end{tabular}

\end{table*}

\begin{table*}[htbp]
\renewcommand\arraystretch{0.5}
\centering
    \caption{\label{tab:table-name}Ablation study on the simulated datasets with various SNRs: basic plus adversarial loss (denoted as Basic$\_$adv), basic plus adversarial loss and cycle-consistency loss method (denoted as Basic$\_$adv$\_$cyc), basic plus adversarial loss and contrastive loss (denoted as Basic$\_$adv$\_$contra) and basic plus adversarial loss, cycle-consistency loss, and contrastive loss (Con-AAE). 
    The performance of Con-AAE is in bold. Clearly, Con-AAE's performance is very stable across different data sizes and SNRs, compared to the other baseline methods.
}
    


	\begin{spacing}{1.15}
	\begin{tabular}{c|c|c|c|c|c|c|c|c|c}
	\toprule[2pt]

	\#Sample	 & Method	& No Noise	& SNR25	& SNR20 & SNR15 & SNR10 & SNR5 & SD & AVG \\
	
	\hline
	\multirow{4}*{1200} & 
	 Basic\_adv    & 80.4 &	83.3 &	83.3 &	81.3 &	82.1 &	76.7 &	2.47 &	81.18 \\
	 & Basic\_adv\_cyc	  & 83.3 &	83.3 &	83.8 &	83.3 &	84.2 &	82.3 &	0.64 &	83.36 \\
	 & Basic\_adv\_contra  & 85.0 &	85.4 &	85.0 &	87.5 &	83.8 &	72.5 &	5.38 &	83.20 \\
	  & \textbf{Con-AAE} &	\textbf{87.5}  & \textbf{86.7} & \textbf{87.9} &	\textbf{88.3} &	\textbf{87.1} &	\textbf{81.7} &	\textbf{2.43} &	\textbf{86.53}\\

	\hline
	\multirow{4}*{2100} & 
	Basic\_adv	 & 85.0 &	85.0 &	84.2 &	82.9 &	84.5 &	81.4 &	1.42 &	83.80 \\ 
	& Basic\_adv\_cyc	& 72.4 &	72.4 &	71.2 &	70.4 &	76.2 &	77.4 &	2.82 &	73.33 \\
	& Basic\_adv\_contra	& 91.0 	& 90.0 & 	91.9 &	91.4 &	86.2 &	72.1 &	7.63 &	87.10 \\
	 & \textbf{Con-AAE} &	\textbf{89.3} &	\textbf{88.8} &	\textbf{88.6} &	\textbf{90.0} &	\textbf{87.8} &	\textbf{82.6} &	\textbf{2.67} & \textbf{87.85} \\

	\hline
	\multirow{4}*{3000} & 
	Basic\_adv	& 81.1 &	82.5 &	82.5 &	81.8 &	77.2 &	75.3 &	3.06 &	80.07 \\
	& Basic\_adv\_cyc &	85.8 &	86.8 &	87.0 &	86.5 &	84.0 &	80.6 &	2.47 &	85.12 \\
	& Basic\_adv\_contra &	88.8 &	90.0 &	89.2 &	67.5 &	60.8 &	86.5 &	12.87& 	80.47 \\
	 & \textbf{Con-AAE} &\textbf{89.7} &\textbf{88.6} &\textbf{88.2} &	\textbf{89.1} &	\textbf{74.2} &	\textbf{82.0} &	\textbf{6.12} &	\textbf{85.30} \\

	\hline
	\multirow{4}*{6000} & 
	Basic\_adv     & 81.0 &81.2 &74.2 & 	67.5 &	44.7 &	35.8 &	19.33 & 64.07\\ 
	& Basic\_adv\_cyc &	81.1 &	85.2 &	85.5 &	78.6 &	79.2 &	78.5 &	3.24 &	81.35 \\
	& Basic\_adv\_contra &	90.8 &	90.9 &	91.3 &	90.0 &	81.9 &	67.6 &	9.43 &	85.42 \\
	 & \textbf{Con-AAE}	& \textbf{87.5}  &	\textbf{89.7} &	\textbf{88.3} &	\textbf{87.3} &	\textbf{82.9} &	\textbf{77.1} &	\textbf{4.69} &	\textbf{85.47} \\
	
	\bottomrule[2pt]
	\end{tabular}
	\end{spacing}

\end{table*}

\end{document}